\documentclass[12pt]{article}
\usepackage{amsmath,amssymb,color}
\usepackage{cite,epsf}

\makeatletter \@addtoreset{equation}{section} \makeatother

\usepackage{graphicx,array} 
\newcommand{\be}{\begin{equation}}
\newcommand{\ee}{\end{equation}}
\newcommand{\beq}{\begin{equation}}
\newcommand{\eeq}{\end{equation}}
\newcommand{\bea}{\begin{eqnarray}}
\newcommand{\eea}{\end{eqnarray}}

\newcommand{\ba}{\begin{eqnarray}}
\newcommand{\ea}{\end{eqnarray}}

\begin{document}

\begin{titlepage}
\vspace{10pt}
\hfill
{\large\bf HU-EP-08/25}
\vspace{20mm} 
\begin{center}

{\Large\bf  Matching gluon scattering amplitudes and\\[2mm] Wilson loops 
in off-shell regularization}

\vspace{40pt}

{\large  Harald Dorn, Charlotte Grosse Wiesmann  }

\vspace{20pt}

{\it Humboldt--Universit\"at zu Berlin, Institut f\"ur Physik\\ Newtonstr. 15, D--12489 Berlin}\\ [4mm]

\vspace{10pt}

\centerline{\tt dorn@physik.hu-berlin.de, lottagw@yahoo.de}

\vspace{40pt}
\end{center}

\centerline{{\bf{Abstract}}}
\vspace{15pt}
\noindent
We construct 
a regularization for light-like polygonal Wilson loops in ${\cal N}=4$ SYM, 
which matches them to the off-shell MHV gluon scattering
amplitudes. Explicit calculations are performed for the 1-loop four gluon
case. The off light cone extrapolation has to be based on the local
supersymmetric Wilson loop. 
The observed matching concerns Feynman gauge. Furthermore, the leading infrared
divergent term is shown to be gauge parameter independent on 1-loop level. 

\end{titlepage}
\newpage
\noindent
{\large\bf Introduction}\\[2mm]
\noindent
The mapping of planar gluon MHV scattering amplitudes in ${\cal N}=4$ SYM
to Wilson loops for polygons built out of light-like segments and their
evaluation via related string configurations at strong 't Hooft coupling
\cite{am} has widened the toolkit of AdS/CFT considerably. Soon after this, it
has been realized that the correspondence between scattering amplitudes
and Wilson loops also holds at weak 't Hooft coupling as a property
of perturbative gauge field theory \cite{k1,b,k2}. Both partners
of the correspondence are divergent. Therefore, one has to specify
regularizations on both sides to 
get a precise statement out of the AdS/CFT lore, which is behind this
correspondence. So far this has been done with dimensional regularization on
both sides, matching its dual IR/UV aspects. 

Given the need to specify a regularization, it is a natural question
to ask how robust the correspondence is with respect to the choice
of regularizations. For scattering amplitudes, besides dimensional
regularization, there is another very natural regularization: going
off shell with the external momenta. In ref. \cite{k1} it has been 
stressed, that this off-shell regularization is sensible to a larger
soft kinematical region than dimensional regularization, resulting in a
characteristic factor 2 in the evolution equation. This observation
gives an additional motivation to find the corresponding regularization
on the Wilson loop side.

The aim of this short note is to identify exactly this wanted regularization
out of the study of the simplest case, the four-gluon amplitude in one loop
approximation. Since the construction will turn out to be very natural, 
we conjecture its validity in general. 
 
The quantity which now has to be matched by the Wilson loop calculation
is the one-loop four-gluon amplitude (divided by the tree
amplitude) with all four momenta slightly off shell, i.e. $p_i^2=-m^2$, 
\cite{k1}
\beq
{\cal M}_4~=~1~+~a~M^{(1)}~+~{\cal O}(a^2)\label{ampl}
\eeq
with
\beq
a~=~\frac{g^2N}{8\pi^2}~\label{a}
\eeq
and ($s,t$ Mandelstam variables)
\beq
M^{(1)}~=~-\log ^2\left (\frac{m^2}{-s} \right )-\log ^2\left
    (\frac{m^2}{-t} \right )~+~\frac{1}{2}\log ^2\frac{s}{t}~-~\frac{\pi^2}{6}~+~{\cal O}(m)~.
\label{M}
\eeq
\\[5mm]
\noindent
{\large\bf The Wilson loop partner for off-shell scattering amplitudes  }
\\[2mm]
A first option could be to keep the identification of the sides of the
tetragon  with the momenta of the scattering amplitude also taken off shell.
While this makes the scattering amplitude finite, the corresponding Wilson
loop is still divergent due to the cusp divergences (now  cusps between
space-like sides). If we handle these by dimensional regularization
$(D=4-2\epsilon )$ and treat the RG scale parameter as in \cite{k1}, i.e. $ g^2\rightarrow g^2 (\mu ^2\pi e^\gamma)^{-\epsilon}$,   we get
\beq
{\cal W}_4~=~1~+~a~W^{(1)}~+~{\cal O}(a^2)   
\eeq
with
\bea
W^{(1)}|_{\mbox{\scriptsize off-cone,
    $\epsilon$}}&=&\frac{1}{\epsilon}\left (\log (\frac{\mu
    ^2}{-s})+\log(\frac{\mu ^2}{-t})+2\log(\frac{m^2}{\mu^2})\right
)+\log^2(\frac{m^2}{\mu^2})-\frac{1}{2}~\log ^2(\frac{\mu^2}{-s})\nonumber\\
&&-\frac{1}{2}~\log ^2(\frac{\mu^2}{-t})
+\frac{1}{2}~\log ^2(\frac{s}{t})+\frac{\pi^2}{2}+{\cal O}(m^2\log m)+{\cal
  O}(\epsilon)~. \label{offandeps}
\eea
Suppressing the pole term in $\epsilon$ yields the renormalized
$W^{(1)}|_{\mbox{\scriptsize off-cone, ren}}$, which depends on  $\mu ^2$ and
$m^2$.  For our purpose it seems natural to
identify $\mu ^2$ and $m^2$, which results in\footnote{That this renormalized
expression diverges for $m^2\rightarrow 0$ is a reflection of the divergence
of the cusp anomalous dimension for space-like sides if the sides approach a
light-like limit.}  
\beq
W^{(1)}|_{\mbox{\scriptsize off-cone, ren}}~=~-\frac{1}{2}~\log
^2(\frac{m^2}{-s})-\frac{1}{2}~\log ^2(\frac{m^2}{-t}) 
+\frac{1}{2}~\log ^2(\frac{s}{t})+\frac{\pi^2}{2}+{\cal O}(m^2\log m)
~.\label{offandren}  
\eeq
Obviously, this does not match (\ref{M}), the main obstacle is the factor
$\frac{1}{2}$ 
in front of the terms quadratic in $\log m^2$. This remains true also
if one keeps $\mu^2$ fixed and looks at $m^2\rightarrow 0$.
As a side remark, note that the difference in the finite
terms independent of the Mandelstam variables, could even be removed by
including a suitable numerical factor in the relation between  $\mu ^2$ and
$m^2$. 

The mismatch cannot be avoided either, by skipping the intermediate
step via the renormalized Wilson loop. Trying this, one immediately would 
have to use in
(\ref{offandeps}) the familiar mapping of dimensional
regularization to  regularizations with dimensional parameter: 
$\frac{1}{\epsilon}\sim -\log(\frac{m^2}{\mu ^2})$.\\

After this unsuccessful attempt, we stay in four dimensions and consider a
combination  of $p^2_i=-m^2$ with a suitable cutoff for the contour parameter
$\tau$. As long as the sides of the polygon are
light-like the scalars decouple. Therefore, one can use either
the Wilson loop for gauge fields $A_{\mu}$ only, or the  local
supersymmetric  Wilson loop coupling to $A_{\mu}$ and the scalars $\phi _I$
\cite{malda, gross}
\beq 
{\cal W}~=~\langle ~ \frac{1}{N}~\mbox{tr P}\exp\left (i\int (A_{\mu}\dot x^{\mu}+\sqrt{\dot
  x^2}~\theta ^I\phi _I)~d\tau\right )~\rangle~.\label{loop}
\eeq   
However, as soon as the sides are no longer light-like, one has to make
a decision. \footnote{Out of  the formal construction according to AdS/CFT in 
\cite{am}, the local supersymmetric version seems to be the natural
partner. Further support arises from the treatment of divergences.} 
We choose the local supersymmetric loop (\ref{loop}), since then
the divergences arising from the limit of coincident propagator end points 
in  the interior of a side of the polygon cancel each other. One only has to
specify a cutoff for the contour parameter near the cusps. 

Cutting out part of the closed contour breaks gauge invariance. But this is
no obstruction, since the partner on the side of the scattering amplitude is 
not gauge invariant off shell either. 
What we now try to match are quantities obtained by
using Feynman gauge on both sides. We will add some more comments on gauge
variance versus invariance below.

We normalize the dimensionless contour parameter $\tau$ by assigning an
interval of length $1$ to each side of the polygon. To relate the cutoff
for the dimensionless $\tau$ to the dimensionful $m^2$, we need another
dimensionful quantity to form a quotient. For the cusp spanned by the
momenta $p_j$ and $p_k$ there is only the scalar product $p_jp_k$
available. Since $\tau$ 
characterizes a portion of momentum, the wanted quotient should be proportional
to $m$ instead of $m^2$. Our choice for the cutoff in $\tau$ at the cusp
spanned by $p_j$ and $p_k$ then is
\beq 
\sigma_{jk}~=~\sqrt{\frac{m^2}{-2p_jp_k}}~.
\label{sigma}
\eeq
The minus sign under the square
root means that we first define our regularized Wilson loop in the kinematical
region, where all scalar products $p_jp_k$ are negative 
\footnote{We have chosen the mostly minus metric.} and then continue from
this region analytically. For the four gluon scattering this region is just the
$u$-channel.  
\begin{figure}[h!]
\begin{center}
\includegraphics[height=5.0cm]{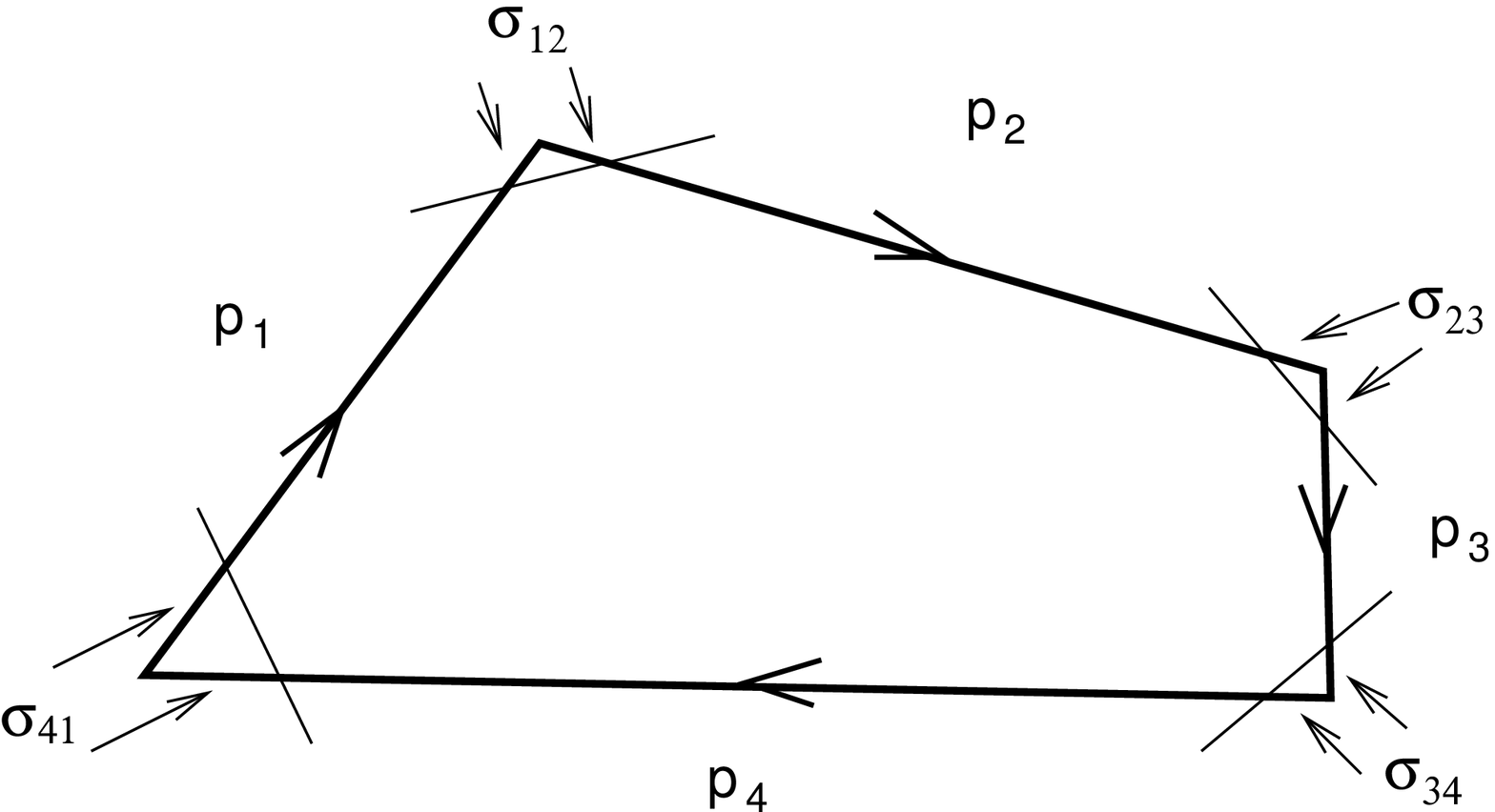}
\\[5mm]
 \end{center}
\noindent
{\bf Fig.1}~~{\it A tetragon contour for the Wilson loop with cut cusps.}
\end{figure}

There is still another motivation for the choice (\ref{sigma}).   
For points $x_j,~x_k$ on the $p_j$-side or $p_k$-side, respectively, let us
first cut out
the region
\beq
 \vert(x_j-x_k)^2\vert\leq m^2~.\label{geomtaucut}
\eeq
Relating the forbidden Minkowski space distances directly to the infrared
regulator of the scattering amplitude, this is a very natural
starting prescription. 
 With the parametrization
$x_j=-\tau_j~p_j,~~x_k=\tau_k~p_k$ and $\tau_j,\tau_k\in [0,1]$, in the
$(\tau_j,\tau_k)$-plane, this
corresponds to cutting the region (note: $p_j^2=p_k^2=-m^2,~p_jp_k<0$)
\beq
\tau_j^2~+~\tau_k^2~+~\frac{(-2p_jp_k)}{m^2}~\tau_j\tau_k~\leq 1
\label{taucut}
\eeq
out of the unit square. The crucial point is that here the quotient
$\frac{(-2p_jp_k)}{m^2}$ appears of its own volition. The hyperbola saturating
the inequality in (\ref{taucut}) has its nearest point relative to the
origin at
$\tau_j=\tau_k=(2+\frac{(-2p_jp_k)}{m^2})^{-1/2}=\sigma_{jk}(1+{\cal O}(\sigma
  _{jk}^2))$. 
Since divergences arise
only when {\it both}, $x_j$ and $x_k$, approach the cusp, eq.(\ref{geomtaucut})
cuts out too much. If we then use the behaviour of the hyperbolas in the
vicinity of the origin to determine the size of the cut squares, we just    
get  (\ref{sigma}) for their length.
In figure 2 we show the relation of the cut regions according to
eqs.(\ref{geomtaucut}) and (\ref{taucut}) in comparison to the above argued
cut $\tau_j,\tau_k\leq 
\sigma_{jk}$ for various values of $\sigma_{jk}$.
\begin{figure}[h!]
\begin{center}
\includegraphics[height=4.5cm]{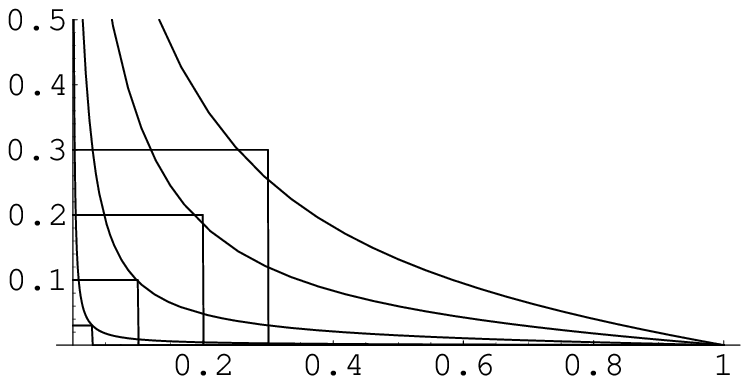}
\\[5mm]
 \end{center}
\noindent
{\bf Fig.2}~~{\it Cut regions in $(\tau_j,\tau_k)$-plane according to 
eqs.(\ref{geomtaucut}) and (\ref{taucut}) compared to $\tau_j,\tau_k\leq
\sigma_{jk}$, for $\sigma_{jk}$ = 0.3, 0.2, 0.1, 0.03, respectively .}
\end{figure}

After these preparations the one loop contribution to the local supersymmetric
Wilson loop for the tetragon of Fig.1 is given by
\beq
W^{(1)}~=~I^{\mbox{\scriptsize cusp}}_{12}~+~I^{\mbox{\scriptsize cusp}}_{23}~+
~I^{\mbox{\scriptsize cusp}}_{34}~+~I^{\mbox{\scriptsize
    cusp}}_{41}~+~I_{24}~+~I_{13}~,\label{w1tot} 
\eeq 
with
\bea
I^{\mbox{\scriptsize
    cusp}}_{jk}&=&\int_{\sigma_{jk}}^1d\tau_j\int_{\sigma_{jk}}^1d\tau_k~
\frac{p_jp_k+m^2}{-(\tau_jp_j+\tau_kp_k)^2+i\varepsilon}\label{Icusp}
\eea
and 
\beq
I_{24}~=~\int_{0}^1d\tau_2\int_{0}^1d\tau_4~
\frac{p_2p_4+m^2}{-(\tau_2p_2-(p_2+p_3+\tau_4p_4))^2+i\varepsilon}~.
\eeq
There is an analogous formula for $I_{13}$. Note that, due to the total
cancellation between the gauge field and the scalar, there is no contribution
to  
(\ref{w1tot}) with both ends of the propagators on the same side of the
polygon. Using polar coordinates we get from (\ref{Icusp}) for $p_jp_k<0$
\bea
I^{\mbox{\scriptsize
    cusp}}_{jk}&=~&2
~(b_{jk}+1)~\int_0^{\pi/4}~\frac{d\varphi}{1-b_{jk}\sin
  2\varphi 
+i\varepsilon}~\int
_{\frac{\sigma_{jk}}{\cos\varphi}}^{\frac{1}{\cos\varphi}}~\frac{dr}{r}
\nonumber\\ 
&=& \log\sigma_{jk}~\frac{b_{jk}+1}{b_{jk}\sqrt{1-1/b_{jk}^2}}~\left (
  \log(-b_{jk})+\log(1+\sqrt{1-1/b_{jk}^2})\right )~,\label{Icusp2}
\eea 
where we introduced $b_{jk}~=~ \frac{p_jp_k}{m^2}$. For a later continuation
to positive $p_jp_k$, one has to make more precise the issue of the phase
of $b_{jk}$. We need the behaviour of $I^{\mbox{\scriptsize cusp}}_{jk}$
for large $b_{jk}$. The only place where this is relevant is in the
term $\log(-b_{jk})$. Due to the $i\varepsilon$ prescription in the first line
of 
(\ref{Icusp2}) this has to be understood as $\log(e^{i\pi}b_{jk})$. 

The limit for $I_{24}+I_{13}$ at $m^2\rightarrow 0$ is finite and can be
taken from \cite{k1,b}. Then, expressing the products $p_jp_k$ in terms of the
Mandelstam variables ($2p_1p_2=2p_3p_4=s+2m^2,~2p_2p_3=2p_1p_4=t+2m^2$), using
(\ref{w1tot}), (\ref{Icusp2}) and the relation of the cutoff to $m^2$ via
(\ref{sigma}), we arrive at
\beq
W^{(1)}~=~-\log ^2\left (\frac{m^2}{-s} \right )-\log ^2\left
    (\frac{m^2}{-t} \right )~+~\frac{1}{2}\log
  ^2\frac{s}{t}~+~\frac{\pi^2}{2}~+~{\cal O}(m^2\log m^2)~.\label{W1fin}
\eeq
Up to a constant independent of $s$ and $t$, this agrees with the off-shell
scattering amplitude (\ref{M}). 

If one alternatively uses an unmodified cutoff according to
(\ref{geomtaucut}), one gets an unwanted factor
$\frac{1}{2}$ in front of the squared logarithms, as in (\ref{offandren}).
It would be interesting to find out, whether this effect is related
to the different soft regions for the scattering amplitude mentioned 
in \cite{k1}. 
\\[5mm]
\noindent
{\large\bf Comments on gauge variance/invariance}
\\[2mm]
As mentioned above, the off-shell scattering amplitude as well as our cut
Wilson loop are not gauge invariant. The observed agreement of both sides
concerns Feynman gauge. However, due to the role of the coefficient in front
of the leading term ($\propto(\log m^2)^2$) in the evolution equation for
the scattering amplitude \cite{k1}, one should expect, that the leading term in
(\ref{W1fin}) is gauge invariant. 

The gauge field propagator in generalized Feynman gauge with gauge parameter
$\alpha$ can be written as
\bea
G_{\mu\nu}^{(\alpha)}(x-y)~=~\frac{\eta_{\mu\nu}}{4\pi
  ^2((x-y)^2-i\varepsilon)}
~+~\frac{\alpha -1}{16\pi
  ^2}~\partial_{\mu}\partial_{\nu}~\log (\Lambda ^2(x-y)^2-i\varepsilon)~.
\label{prop}
\eea 
The auxiliary scale parameter $\Lambda ^2$ drops out after performing
the differentiation. Its only purpose is to start with the logarithm
of a dimensionless quantity. The propagator of the scalar field is independent
of $\alpha $. Therefore, the gauge dependent term to the one loop contribution
$W^{(1)}$ for a generic closed contour is
\beq
W_{\mbox{\scriptsize gauge}}^{(1)}~=~\frac{\alpha -1}{16\pi ^2}
\int_{\tau_1>\tau_2}~ d\tau_1d\tau _2~\frac{d}{d\tau_1}\frac{d}{d\tau_2}~\log (\Lambda
^2(x(\tau_1)-x(\tau_2))^2-i\varepsilon)~.\label{Wgauge} 
\eeq 
Due to the $i\varepsilon$ prescription this term needs no regularization, it
is identically  zero before $\varepsilon$ is sent to zero. This situation 
changes for
our loop cut according to fig.1. In addition,  the limit
$\varepsilon\rightarrow 0$  only exists, if before, an additional cutoff
$\sigma$ 
is introduced, preventing the coincidence of $\tau_1$ and $\tau_2$ on the same
side of the polygon. For this it would be natural to choose $\sigma =\Lambda
m$. Thus, to handle generalized Feynman gauge we pick
up another scale parameter $\Lambda$ for dimensional reasons.  But in any
case,  
due to the total derivative structure in 
(\ref{Wgauge}), the final result for the gauge dependent contribution
can contain only $\log$-terms and no $\log ^2$-terms. This proves the gauge
independence of the leading $\log^2$-term in our one loop result of the
previous paragraph. 
\newpage
\noindent
{\large\bf Conclusions}
\\[2mm]
We have given various arguments for a regularization of polygonal Wilson
loops that exactly matches the MHV off-shell gluon amplitudes. This is true 
up to finite
terms independent of the Mandelstam variables and terms vanishing for
$m^2\rightarrow 0$.  While for 
light-like polygons there is no difference between the pure gauge field 
Wilson loop and the local supersymmetric version, our off shell extension
is based on the supersymmetric loop. The proposal has been 
checked explicitly on 1-loop level for the four gluon case.  Furthermore, it 
has been shown,
that the leading $\log^2$ contribution is independent of the gauge parameter
of generalized Feynman gauge. 

An obvious continuation of this work should be the inclusion of higher
orders in perturbation theory. Our regularization could also be of interest in
the discussion of the 
dual conformal symmetry of scattering amplitudes and Wilson loops 
\cite{k1,k2,k3}.

Turning to the strong coupling situation, one should construct a regularization
for the string approaching an off light cone polygon at the boundary of AdS.
There an alternative to dimensional regularization, but still for the
light-like case, has already been discussed in \cite{alday}. It uses a cutoff
in the radial direction which is position dependent similar to our
construction.
\\[15mm]
{\bf Acknowledgement}\\[2mm] 
\noindent
This work has been supported in part by the German Science Foundation (DFG)
grant DO 447/4-1. We thank Nadav Drukker, George Jorjadze, Jan Plefka, Arkady
Tseytlin and Donovan Young for useful discussions and Gregory Korchemsky for
a helpful e-mail.

\end{document}